\documentclass[11pt]{article}

\usepackage[usenames,dvipsnames,table]{xcolor}

\usepackage{jheppub} 

\usepackage{graphicx}
\usepackage{amsmath, amssymb}
\usepackage{youngtab}
\usepackage{manfnt}
\usepackage{float}
\usepackage{placeins}

\newcommand{\anti}{\raisebox{-1.1ex}{\epsfxsize=0.12in\epsfbox{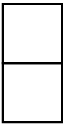}}}
\newcommand{\sym}{\raisebox{-0.2ex}{\epsfxsize=0.22in\epsfbox{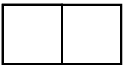}}}

\newcommand{\be}{\begin{eqnarray}}
\newcommand{\ee}{\end{eqnarray}}
\def\eqn{\eqref}

\newcommand{\nn}{\nonumber}

\title{Comments on Symmetric Mass Generation in 2d and 4d}

\author{David Tong}

\affiliation{Department of Applied Mathematics and Theoretical Physics \\ University Cambridge, CB3 0WA, UK}

\emailAdd{d.tong@damtp.cam.ac.uk}

\abstract{Symmetric mass generation is the name given to a  mechanism for gapping fermions while preserving a chiral, but necessarily non-anomalous, symmetry.  In this paper we describe how symmetric mass generation for continuous symmetries can be achieved using gauge dynamics in two and four dimensions. Various strong coupling effects are invoked, including  known properties of  supersymmetric gauge theories, specifically the phenomenon of s-confinement, and conjectured properties of non-supersymmetric chiral gauge theories.}

\begin{document} 

\maketitle
\flushbottom

\section{Introduction}

Global symmetries and their attendant 't Hooft anomalies provide a classification scheme for quantum field theories. These offer a powerful organisational principle that places strong restrictions on dynamics. In particular, theories with different 't Hooft anomalies  can be related by neither RG flow \cite{thooftanomaly} nor duality \cite{nati1,nati2}.

This classification scheme invites a basic question. Given two theories, say A and B,   that exhibit identical 't Hooft anomalies, is it possible to deform one theory into the other? 

To address this, we should first describe what constitutes an allowed deformation of a theory. The basic idea is that one is always free to  introduce new degrees of freedom at some high energy scale. These new degrees of freedom should be consistent with the symmetry that underpins the the classification, and leave both the 't Hooft anomalies and the low-energy physics of the theory unchanged. 

Suppose that one starts with a quantum field theory -- call it Theory A --  together with some new degrees of freedom at some high scale. One deforms Theory A by varying parameters, ensuring that the symmetry remains unbroken. In doing so, these heavy degrees of freedom can come down to a lower scale where they interact with Theory A, changing its dynamics but not its anomalies.  The goal is to find a path in parameter space such that the end point consists of   Theory B at low energies,  typically with a different set of degrees of freedom decoupled at some high energy scale.

By construction, any two theories that are connected by such a path share the same 't Hooft anomalies, including any possible subtle anomalies that are still to be discovered. But is the converse true? If two theories share the same 't Hooft anomalies, can we always find  a path between them? Seiberg has conjectured  that the answer to this question is yes \cite{seibergtalk}.

The purpose of this paper, which is a continuation of our earlier work \cite{shlomo},  is to explicitly construct such paths for theories that lie in the trivial class. These theories exhibit a symmetry $G$ for which the 't Hooft anomalies vanishes. This means that there is, in principle, no obstacle to gapping the theories while preserving the symmetry, leaving us with nothing at low energies. However, although the class is trivial, the path to demonstrating this may not be. A clear example arises when the theory includes fermions transforming in some chiral, but anomaly free representation of $G$. In this case, simple, quadratic mass terms for the fermions are forbidden since they necessarily break the symmetry. If we wish to gap the system, we must be more creative. Mechanisms that successfully  gap theories  without breaking chiral symmetries sometimes  go by the name of {\it symmetric mass generation}.

In recent years there has been impressive progress in understanding symmetric mass generation in various systems, both with underlying discrete and continuous symmetries. In the case of discrete symmetries, examples of symmetric mass generation gave early indications that non-interacting topological states of matter may fail to survive the introduction of interactions \cite{fk,ryuz,qi,fcv,senthil,bentov,max}. The classification was subsequently understood \cite{kap} in terms of the  underlying anomalies \cite{df}.

For continuous symmetries, there is a long history of attempts at symmetric mass generation in the context of lattice gauge theory. In that setting, the goal is to find a way to gap the fermion doublers, leaving the original fermions unscathed, in the hope of finding a route to constructing chiral gauge theories on the lattice. This idea was first proposed by Eichten and Preskill \cite{ep} and interesting variants on the theme have been explored by a number of different authors over the years \cite{creutz,poppitz,wen,request,demarco,kik,ww1,ww2}. Nearly all of these approaches  rely on strong coupling effects at the lattice scale, where irrelevant operators can drive the low energy dynamics. (An important exception to this statement will be highlighted in Section \ref{2dsec}.) However it is difficult to retain control over these theories and finding the phase exhibiting symmetric mass generation is likely to require fine tuning, possibly in a large dimensional parameter space, making it challenging in numerical simulations \cite{golt,poppitz2}. More recently, however,  positive results along these lines have been  exhibited in the context of K\"ahler-Dirac fermions \cite{simon,simon1,simon2}.

\subsubsection*{Symmetric Mass Generation in Continuum Field Theory}

In this paper we describe a number of different mechanisms for symmetric mass generation   in continuum field theory. We will focus on theories in $d=3+1$ and $d=1+1$ dimensions and restrict ourselves to continuous symmetries.

In some ways, the approach in the continuum is similar to that on the lattice: both start by introducing what appear to be  irrelevant (or sometimes marginally irrelevant) operators. Typically, of course, irrelevant operators cannot gap a system. On the lattice one attempts to circumvent this by invoking strong coupling effects at the UV cut-off. In continuum field theory, this is not an option we wish to rely on: all interactions should be weak at the cut-off, but may flow to strong coupling in the infra-red. Such an RG flow can render an irrelevant operator {\it dangerously irrelevant}. This was the mechanism for symmetric mass generation invoked in \cite{shlomo}, and previously promoted in \cite{you1,you2}. The challenge, of course, is to keep enough control over the strong coupling regime so that one can be sure which operators are dangerously irrelevant and that no symmetry is broken en route.

We will show that in several examples  the requisite level of control can be achieved simply by applying known results about strong coupling effects in quantum field theory. In particular, the punchline of \cite{shlomo} was that symmetric mass generation can be induced by any theory that exhibits confinement without chiral symmetry breaking. In Section \ref{4dsec} we describe a number of 4d theories that are known (or conjectured) to have this property and explain how they can be leveraged to achieve symmetric mass generation

Along the way, we also address a number of subtleties. In particular, the observation that, in many theories, the Higgs and confining phases are continuously connected suggests that there may be a weakly coupled description of the symmetrically gapped phase\footnote{I'm grateful to Nathan Seiberg for raising this point.}. As we explain, although this is indeed true it does not mean that there is a path to symmetric mass generation that proceeds entirely through weak coupling: in the examples we give, one must always pass through a strongly coupled point somewhere along the way.

In Section \ref{2dsec}, we turn to symmetric mass generation in $d=1+1$ dimensional field theories. Here the challenges are slightly different, not least because it's not possible to spontaneously break continuous symmetries in two dimensions \cite{mw,coleman}. As we will review, in contrast to the 4d story, it is often possible to implement symmetric mass generation in 2d without the need to first introduce extra, heavy degrees of freedom. Nonetheless, we will see that it is useful to also understand how symmetric mass generation proceeds when implemented by  confining gauge fields. This has the advantage of highlighting the similarities and differences with the 4d examples.

\section{Gapping Chiral Fermions in Four Dimensions}\label{4dsec}

In this section we describe a number of examples of  4d fermions carrying a chiral, non-anomalous representation under a symmetry group $G$. The goal of symmetric mass generation is to gap the fermions, without breaking $G$.

For pedagogical purposes, it is simplest to think of $G$ as a global symmetry, if only because the discussion of broken global symmetries is much cleaner than that of broken gauge symmetries.  However, among our examples is the Standard Model with $G=SU(3)\times SU(2)\times U(1)$ and it seems perverse to insist that $G$ is viewed as a global symmetry in that case. Furthermore, in all examples the symmetry group $G$ will be non-anomalous and so there is no obstacle to promoting it to a gauge symmetry if desired.

\subsubsection*{A Mechanism for Symmetric Mass Generation}

In all examples in this section, symmetric mass generation is achieved in the same way.  Our starting point is a collection of massless, chiral fermions. This is Theory A in the language of the introduction.

We then add a collection of  auxiliary scalars and fermions which, at least initially, we take to be much heavier than any other scale. Because the fermions are heavy, they cannot change the anomaly of $G$. In what follows, we will take these additional fermions to lie in vector-like representations of $G$.

The second step is to identify a new symmetry group $H$ which acts on some subset of the original fermions and, possibly, on some of the additional matter fields too. We require that $H$ is non-anomalous and, moreover, that $[H,G]=0$ and $H$ has no mixed anomalies with $G$. This ensures that we can gauge $H$ without breaking $G$. We further introduce scalar fields charged under $H$, but neutral under $G$, which we can condense to make all $H$ gauge bosons heavy. We now have the massless fermions of Theory A, together with a collection of heavy auxiliary fields.

The next step is to change the parameters so that the auxiliary fields become light and interact with our original fermions. We want to design these extra degrees of freedom so that one can find a $G$-preserving path in which all degrees of freedom become massive. This is desired Theory B: an empty theory in the infra-red.

As shown in \cite{shlomo}, this can be achieved by orchestrating a  situation in which the $H$ gauge theory confines, but without breaking chiral symmetry. In the present context, the chiral symmetry is identified with the symmetry group $G$ that we are striving to protect. To construct a mechanism for symmetric mass generation, one must find a choice of $H$ that binds together  some subset of the original fermions into massless composites, while leaving others untouched, in such a way that the final collection of fermions sits in a vector-like representation of $G$.  The  introduction of some dangerously irrelevant operators in the UV then allows the original collection of fermions to be gapped while preserving $G$.  Details of this construction in specific examples will be given below.

The phenomenon of confinement without chiral symmetry breaking is a striking strong coupling effect. It is known to occur in certain supersymmetric theories, where it goes by the name of s-confinement. Confinement without chiral symmetry breaking  is also conjectured to occur in certain non-supersymmetric chiral theories. However, it is not a phenomenon that we have a good handle on beyond these rather special cases and, as we review below, there are powerful theorems that suggest that it does not happen in many other examples where naive 't Hooft anomaly matching may suggest otherwise.  In the rest of this section, we detail a number of examples of this basic idea.

\newpage

\subsection{$G=SU(N)$ with a Symmetric and Fundamentals}\label{herewego}

 Our first example consists of a well known collection of anomaly free, massless Weyl fermions transforming under
\be G = SU(N)\ \mbox{with  $\sym$ and $(N+4)$  $\overline{\Box}$}\nn\ee
Unless otherwise stated, we will take all fermions to be right-handed. The goal is to find a path to gapping these fermions without breaking $G$. 

As described above, the methods we use revolve around identifying an auxiliary symmetry $H$. In the present case, we do not have to look very far: it is
\be H = SU(N+4)\nn\ee
As it stands,  $H$ is anomalous. We can easily rectify this by introducing an additional collection of  $\frac{1}{2}(N+4)(N+3)$ Weyl fermions which are singlets under $G$ but we will take them to transform in the $\anti$ of $H$. This ensures that $H$ is anomaly free. Because these additional fermions are $G$ singlets, there is no obstacle to giving them a large mass preserving $G$, albeit at the expense of $H$.

The theory also enjoys an additional $U(1)$ symmetry that has no mixed anomaly with $G\times H$. This won't play a large role in what follows, but we include the quantum numbers for completeness. The upshot is a collection of fermions transforming under $G\times H\times U(1)$ which we denote as

\FloatBarrier
\begin{table}[h!]
\begin{center}
\begin{tabular}{c|c|c|c}
& $ \lambda$ & $\psi$ & $\chi$  \\
  \hline 
   $G$ & $\sym$ & $\overline{\Box}$  & ${\bf 1}$ \\
  $H$ & ${\bf 1}$ & $\overline{\Box}$  & $\anti$ \\
   $U(1)$ & $-(N+4)$ & $N+2$  & $-N$ \\    \end{tabular}
 \end{center}
\end{table}
\FloatBarrier

%
%
%
%

\noindent
We now describe two different symmetric gapping mechanisms that give all fermions a mass preserving $G$. Both paths involve strong coupling effects but, as we explain, there is also a weakly coupled description of the final gapped state.

 \subsubsection*{A Supersymmetric Mechanism}

Supersymmetry provides a useful level of control over the strong coupling dynamics of strongly interacting gauge theories. The idea that one could apply  known results about supersymmetric dynamics to better understand topological phases was also promoted in \cite{tachyon1}. As stressed in \cite{shlomo}, in certain circumstances this can be leveraged to carve a path to symmetric mass generation. 

For the present example, we start by  gauging the subgroup 
\be SO(N+4) \subset H\nn\ee
In addition, we must introduce a number of scalar fields. Their role is twofold. First, we introduce scalar fields charged under $H$, but neutral under $G$, which, when condensed, give masses to the gauge fields $H$ and the additional fermions $\chi$. In this phase, we are left only with the original massless fermions $\lambda$ and $\psi$ and, of course, $G$ unbroken.

Next, we introduce further scalar fields that have the same quantum numbers under $G\times H$ as $\lambda$ and $\psi$. We call these scalars $M$ and $\phi$ respectively. It is useful to require the  $U(1)$ quantum numbers to differ from the fermions, so  we have
\FloatBarrier
\begin{table}[h!]
\begin{center}
\begin{tabular}{c|c|c}
& $M$ & $\phi$   \\
  \hline 
   $G$ & $\sym$ & $\overline{\Box}$   \\
  $SO(N+4)$ & ${\bf 1}$ & ${\Box}$  \\
   $U(1)$ & $-4$ & $2$   \\    \end{tabular}
 \end{center}
\end{table}
\FloatBarrier

\noindent
For $SO(N+4)$, of course, there is no distinction between $\Box$ and $\overline{\Box}$. 
Originally these scalars are  also given a large mass to decouple them from the dynamics. We also endow them with a Yukawa coupling 
\be {\cal L}_{\rm Yuk} = \sqrt{2} \phi^\dagger \chi \psi  + {\rm h.c.}\label{yuk1}\ee
This will only affect the dynamics as the scalar $\phi$ becomes light. 

Now we specify the path through theory space. We vary the potential of the first set of scalar fields so that they uncondense and then vary it further so that they become heavy and decouple. Consequentially, both the $H$ gauge bosons and the $\chi$ fermion become light. At the same time, we dial the potential of $\phi$ until it becomes massless and the theory exhibits ${\cal N}=1$ supersymmetry.  The Yukawa coupling \eqn{yuk1} has the value required for a supersymmetric theory.

The final theory is $H= SO(N+4)$ SQCD, coupled to $N$ chiral multiplets $\Phi$ in the fundamental representation. The group $G=SU(N)$ arises as a global chiral symmetry group for this theory. The chiral multiplets $\Phi$ contain $\phi$ and $\psi$, while $\chi$ acts as the gaugino. The additional $U(1)$ symmetry shown in the tables above coincides with the R-symmetry of the supersymmetric theory\footnote{Usually the R-symmetry is normalised so that the gaugino $\chi$ has charge 1, in which case other fields typically have fractional charge. Here charges are rescaled by a factor of $N$.}.  For now, the scalar $M$ and fermion $\lambda$  -- which live in a chiral multiplet that we call ${\cal M}$ -- are decoupled from the dynamics.

The benefit of tuning to the ${\cal N}=1$ supersymmetric point is that we have a good handle on the strong coupling dynamics. In the present case, this was determined by Intriligator and Seiberg \cite{is} who showed that, at the origin of moduli space, this theory has the special property of {\it s-confinement}, meaning that it confines without breaking the $G=SU(N)$ chiral symmetry. The low-energy theory consists of ${\cal N}=1$ massless composite fields which, in terms of superfields, are 
\be \widetilde{\cal M} = \Phi\Phi\nn\ee
with the fields contracted so that $\widetilde{\cal M}$ is a singlet under the $H=SO(N+4)$  gauge symmetry. It then transforms in the $\overline{\sym}$ representation of $G=SU(N)$. 

Crucially, the  composite fermions transform in the conjugate representation to the superfield ${\cal M}$, which contains the scalars $M$ and the fermions $\lambda$ that were not coupled to  $H$. This means that it is now trivial to gap the system without breaking the  $G=SU(N)$ global symmetry. We simply add a superpotential in the UV
\be {\cal W}_{UV} = y\,{\cal M}\Phi\Phi\nn\ee
invariant under $G\times H\times U(1)$. Here $y$ is a dimensionless Yukawa coupling. Naively this superpotential is marginally irrelevant. However the strong coupling dynamics means that it is, in fact, dangerously irrelevant and, after confinement, the UV superpotential descends to a mass term
\be {\cal W}_{IR} = y \Lambda\,{\cal M}\widetilde{\cal M}\nn\ee
giving all fields a mass $y \Lambda$, where $\Lambda$ is the strong coupling scale of the $SO(N+4)$ gauge theory.

As this example shows, the phenomenon of confinement without chiral symmetry breaking can successfully transmute a chiral, but anomaly free representation of $G$ into a vectorlike representation, whereupon it is trivial to gap the system. 

The role of supersymmetry above was to provide a crutch in the analysis, providing precious information about the strong coupling dynamics. However, importantly, the resulting symmetric gapped phase is robust since the spontaneous breaking of $G$ would necessarily be accompanied by gapless Goldstone modes, and these cannot arise from arbitrarily small deformations of a gapped phase. This ensures that the symmetric gapped phase is, as the name suggests, a phase rather than a point: it persists in some neighbourhood the supersymmetric point \cite{shlomo}.

\subsubsection*{A Putative Non-Supersymmetric Mechanism:}

We now return to our starting point of free, massless fermions and  try to find a different mechanism that does not rely on supersymmetry. A good candidate is to simply gauge $H=SU(N+4)$. 

In this case we must again add scalars which, upon getting an expectation value, succeed in decoupling the $H$ gauge bosons.  However, this time, in order to affect symmetric mass generation we need only give these additional scalars a large positive mass and remove them from the interesting dynamics. We are then left with the strong coupling dynamics of the chiral gauge theory
\be \mbox{$H=SU(N+4)$ with $\lambda$ in the $\anti$ and $N$ fermions $\psi$ in the $\overline{\Box}$}
\label{h1}\ee
This is a famous chiral gauge theory, first studied in \cite{raby}. The low-energy dynamics is not fully understood, but a good candidate is confinement without chiral symmetry breaking. In this scenario, the 't Hooft anomalies in the $G\times U(1)$ global symmetry are matched by the massless composite
\be \tilde{\lambda} = \psi\chi \psi\label{mcomposite}\ee
which, once again, transforms in the  $\overline{\sym}_{N+4}$ representation of $G\times U(1)$.

If this dynamics indeed holds, then  a symmetric, gapped phase is attained by the addition of the $G\times U(1)$ invariant 4-fermion term
\be {\cal L}_{\rm 4-fermi} \sim  \frac{\psi\chi\psi\lambda}{M_{UV}^2} + {\rm h.c.} \label{4fermi}\ee
where $M_{UV}$ is a high, UV scale.
This is a dangerously irrelevant operator in the UV and, assuming confinement without chiral symmetry breaking, flows to a simple mass term $\tilde{\lambda}\lambda$ in the infra-red. This time, the mass is of order ${\cal O}(\Lambda^3/M_{UV}^2)$, with $\Lambda$ the strong coupling scale. 

Before proceeding, it is worth making a few comments on the current status of this proposed confining phase. The original argument for confinement without chiral symmetry breaking was simply that $\tilde{\lambda}$ offers a particularly natural candidate for the massless composite fermion, while one has to look to relatively high dimension operators to find a gauge invariant order parameter for chiral symmetry breaking.  Furthermore, a rather different path leads to the same conclusion: if one assumes that there is a Higgs mechanism, with a condensate for a (necessarily gauge dependent) fermion bilinear then, with some natural assumptions, the  global symmetry is unchanged, albeit after  after a twist with the gauge group, and one finds a massless spectrum with the same quantum numbers in the infra-red \cite{raby}.  (We'll review the physics of this  Higgs phase shortly) The lack of Nambu-Goldstone bosons from chiral symmetry breaking was buttressed in \cite{zepp} by a large  $N$ argument, although this has recently been disputed \cite{ken}.

Nonetheless, some doubts have been raised about the proposed phase with the massless composite fermion \eqn{mcomposite} and unbroken global symmetry. In \cite{apple} is was suggested  that a six-fermion condensate could form, breaking the global symmetry. They point out that the resulting massless Goldstone bosons have a lower free energy than the massless fermion \eqn{mcomposite}, although there is no strong reason that this should be invoked as a diagnostic for the preferred low-energy phase\footnote{More recently, the authors of \cite{stefano1,stefano2} made a detailed study of the symmetry group of this theory, including certain discrete quotients. They claim that the theory has a novel mod 2 anomaly that cannot be realised in the free fermion confining phase but is allowed in the Higgs phase. However, this anomaly appears to be ficticious, arising due to an incorrect normalisation of the ${\bf Z}_2$ gauge field \cite{unpublished}.}.

\subsubsection*{The Gapped Phase At Weak Coupling}

It was pointed out, long ago,  that the low-energy confining phase, with massless fermion $\tilde{\lambda} = \psi\chi\psi$, coincides with the Higgs phase  of the theory \cite{raby}.

The original argument invoked strong coupling, with the gauge symmetry broken by a fermion bilinear. However, it is more straightforward to make the same argument using weak coupling physics \cite{seibergtalk}.  To this end, we once again gauge $H=SU(N+4)$. We further introduce a collection of  scalar fields $\phi$ transforming as $(\overline{\Box},{\Box})_{+2}$ of $G\times H\times U(1)$ and then introduce the Yukawa interactions
\be {\cal L} \sim \phi^\dagger\chi\psi + \phi \lambda\psi + {\rm h.c.}\label{yuk2}\ee
Note that the first term coincides with the supersymmetric Yukawa coupling \eqn{yuk1}, except here the gauge group is $SU(N+4)$ instead of $SO(N+4)$.

A potential for  $\phi$ is then constructed so that the expectation value takes the ``colour-flavour" locked form
\be \phi^a_{\ i} = v\,\delta^a_{\ i}\nn\ee
where $a=1,\ldots, N$ is the $G=SU(N)$ index and $i=1,\ldots, N+4$ is the $H=SU(N+4)$ index. Manifestly, the choice of expectation value above means that $\phi^a_{\ i}=0$ when $i=N+1,N+2,N+3$ or $N+4$.

In the colour-flavour locked phase, the $G=SU(N)$ global symmetry survives, but only after twisting by $H$. The symmetry breaking pattern is
\be SU(N) \times SU(N+4)\times U(1) \ \longrightarrow\ SU(N)_{\rm diag} \times SU(4)_{\rm gauge} \times U(1)'\ \ \ \ \label{sym1}\ee
Here $SU(N)_{\rm diag}\times U(1)'$ are global symmetries while, as the name suggests $SU(4)_{\rm gauge}$ is a gauge symmetry. The quantum numbers of the various fermions decompose as \cite{raby}
\be \psi:\ \ & \ \ (\overline{\Box},\overline{\Box})_{N+2}\ &\longrightarrow\ \  \ (\overline{\sym},{\bf 1})_{N+4}\ \oplus\ (\overline{\anti},{\bf 1})_{N+4}\ \oplus\ (\overline{\Box},\overline{\bf 4})_{2+N/2} \nn\\ 
\chi:\ \ &\ \ \ ({\bf 1},\anti)_{-N} \ \ &\longrightarrow\ \ \ (\anti,{\bf 1})_{-N-4} \ \oplus\ (\Box,{\bf 4})_{-2-N/2} \ \oplus\ ({\bf 1},{\bf 6})_0\nn\\ 
{\lambda}:\ \  &\ (\sym,{\bf 1})_{-(N+4)}\  &\longrightarrow \ \ \  (\sym,{\bf 1})_{-(N+4)}\nn\ee
This example again provides a simple illustration of the main point. The quantum numbers of the fermions have transmuted, morphing from a  chiral  representation of the global symmetry $G=SU(N)\times U(1)$ into a vector-like representation of the surviving $SU(N)_{\rm diag}\times U(1)'$. This is achieved by twisting with the gauge group and is possible only because the original, chiral representation was anomaly free.

The Yukawa couplings \eqn{yuk2} now succeed in gapping almost all fermions, leaving $SU(N)_{\rm diag}\times U(1)'$ intact. The only exception is the fermion transforming in the $({\bf 1},{\bf 6})_0$.  But this too becomes gapped once the surviving $SU(4)_{\rm gauge}$ becomes strong and confines. The symmetry $SU(N)_{\rm diag} \times U(1)'$ is gauge equivalent to the original $G\times U(1)$, so succeeds in our original goal of gapping the fermions preserving the chiral symmetry. 

The Higgs mechanism described above provides a weakly coupled description of the gapped phase. However, this does not mean that there is a path from the massless, chiral fermions to the gapped phase that remains weakly coupled throughout. Recall, that the starting point is to add heavy degrees of freedom that leave  $G$ untouched. In the present context, these heavy degrees of freedom are the $H=SU(N+4)$ gauge bosons and these must be made heavy by an expectation value for some scalar that is {\it not} charged under the $G=SU(N)$ chiral symmetry. 

Suppose, for example, that we tried to evade this and instead introduce the $H$ gauge bosons, made heavy through an expectation value for the  $\phi$ scalar, with quantum numbers $(\overline{\Box},\Box)_{+2}$ under $G\times H\times U(1)$. In this case our massless fermions don't carry chiral quantum numbers under $G=SU(N)$ but instead the vectorlike quantum numbers under the unbroken $G=SU(N)_{\rm diag}$. That is not the starting point that we set ourselves!

This means that, in addition to $\phi$, we must have another scalar -- let us call it $h$ -- which carries quantum numbers only under $H$. This can be condensed to make $H$ gauge bosons heavy. We then uncondense $h$ and decouple it from the dynamics and, subsequently, condense $\phi$. Importantly, however, we cannot have both $h$ and $\phi$ condensed at the same time, for this would break our prized $G =SU(N)$ symmetry that we are trying to preserve. This means that to affect symmetric mass generation in this model, our path must pass through the strong coupling point in which neither $h$ nor $\phi$ have an expectation value, and the $H=SU(N+4)$ gauge symmetry is unbroken.

\subsection{$G=SU(N)$ with an Anti-Symmetric and Fundamentals}

The next example consists of another well known collection of massless Weyl fermions, this time  transforming under the global symmetry
\be G = SU(N)\ \mbox{with  $\anti$ and $(N-4)$  $\overline{\Box}$}\label{g1}\ee
with $N\geq 5$. As we shall see, much of the discussion proceeds in parallel with our previous example, and we shall be brief where there is large overlap. However, there are also a number of further subtleties that arise.

Once again,  we do not have to look far to identify an auxiliary symmetry group: 
\be H=SU(N-4)\nn\ee
We can render $H$ non-anomalous by  introducing an additional  $\frac{1}{2}(N-4)(N-3)$ Weyl fermions which are singlets under $G$, but transform in the  $\sym$ representation of $H$.  Including an additional $U(1)$ global symmetry, the quantum numbers of our fermions are

\FloatBarrier
\begin{table}[h!]
\begin{center}
\begin{tabular}{c|c|c|c}
& $ \chi$ & $\psi$ & $\lambda$  \\
  \hline 
 $G$ & $\anti$ & $\overline{\Box}$  & ${\bf 1}$ \\
  $H$ & ${\bf 1}$ & $\overline{\Box}$  & $\sym$ \\
   $U(1)$ & $-(N-4)$ & $N-2$  & $-N$ \\    \end{tabular}
 \end{center}
\end{table}
\FloatBarrier

%
%

\noindent
Clearly this spectrum is similar to the previous example; the difference is that the  role of the global symmetry $G$ and soon-to-be gauge symmetry $H$ have been  exchanged

As we will now see, this means that we must take a slightly different approach to find a symmetric gapped phase since, in contrast to the previous example, the dynamics of the $H=SU(N-4)$ gauge theory, whether strongly or weakly coupled,  may not be sufficient to gap the fermions. (For example,  it is clearly insufficient for the case $N=5$!) This means that we will find ourselves having to add additional vectorlike matter to enhance the group $H$ before turning to its dynamics. 

We now follow the route taken in the previous example and describe different ways to affect symmetric mass generation. 

\subsubsection*{Again, A Supersymmetric Mechanism}

A method to invoke supersymmetric gauge dynamics to give rise to symmetric mass generation was described in \cite{shlomo}. Here we give a brief overview.

The story  is simplest  in the case that $N$ is even, where we gauge
\be Sp(N/2 - 2) \subset H\nn\ee
As in the previous example, we include two sets of scalars. The first are used to Higgs the gauge field $H$ so that our starting point is simply free fermions.
The second set of scalars, which we again call $M$ and $\phi$, are superpartners for $\chi$ and $\psi$. They have quantum numbers

\FloatBarrier
\begin{table}[h!]
\begin{center}
\begin{tabular}{c|c|c}
& $M$ & $\phi$   \\
  \hline 
   $G$ & $\sym$ & $\overline{\Box}$   \\
  $Sp(N/2-2)$ & ${\bf 1}$ & ${\Box}$  \\
   $U(1)$ & $+4$ & $-2$   \\    \end{tabular}
 \end{center}
\end{table}
\FloatBarrier
\noindent
As before, these scalars are originally massive but, as we take the path to the symmetric massive phase, they are tuned to the ${\cal N}=1$ supersymmetric point, including Yukawa interactions of the form
\be {\cal L}_{\rm Yuk} = \sqrt{2} \phi^\dagger \lambda \psi  + {\rm h.c.}\label{yuk4}\ee
The fermions $\lambda$ are the gauginos at the supersymmetric point.

We now  have an ${\cal N}=1$ supersymmetric $Sp(N/2-2)$ gauge theory coupled to $N$ fundamental chiral multiplets. This is again known to exhibit confinement without breaking the $G\times U(1)$ global chiral  symmetry \cite{ip}. The fields once again bind into singlets under $H$, with the massless meson fields. These massless composites have conjugate quantum numbers to $M$ and $\chi$.  A Yukawa term in the UV will once again serve to gap these fermions in the IR.

For $N$ odd, one must work a little harder, adding further additional fermions with vector-like quantum numbers under $G$. Details can be found in \cite{shlomo}.

\subsubsection*{Again, A Putative Non-Supersymmetric Mechanism}

As in the previous case, there is a possible mechanism for gapping at strong coupling without invoking supersymmetry. To achieve this for $N\geq 7$, we need merely gauge $H=SU(N-4)$.
\be \mbox{$H=SU(N-4)$ with $\chi$ in the $\sym$ and $N$ fermions $\psi$ in the $\overline{\Box}$}
\label{h2}\ee
This gauge theory is often discussed alongside \eqn{h1}. As in the previous case, the low-energy dynamics is not well understood but there is a simple conjectural phase in which the theory undergoes confinement without breaking its global $SU(N)\times U(1)$ chiral symmetry. If such dynamics were to take place, it would be accompanied by a massless composite fermion
\be \tilde{\chi} = \psi\lambda \psi\nn\ee
transforming in the $\overline{\anti}_{N-4}$ representation of $G\times U(1)$. A 4-fermion term of the form \eqn{4fermi} again gaps the system.

We stress that we do not know with any level of certainty if this confining phase is indeed realised and the comments made about chiral gauge dynamics in Section \ref{herewego} also pertain here.

The mechanism above clearly fails for $N=5$ since it relies on  $H=SU(N-4)$ strongly interacting gauge dynamics. However, it also fails for a more subtle reason when $N=6$. In this case, the auxiliary gauge dynamics \eqn{h2} involves an $H=SU(2)$ gauge group. Because all representations are either real or pseudo-real, this is a vectorlike theory and this has consequence. It means that, after integrating out the fermions, the path integral measure is positive definite and this, in turn, allows Weingarten-type inequalities to be invoked for the masses of composite states \cite{wein}.  These powerful inequalities can be used to show that the composite scalar $\psi\psi$ is never heavier than the composite fermion $\tilde{\chi} = \psi\lambda\psi$. (A review of the Weingarten inequalities can be found in  Section 5.6.3 of \cite{gt}. An argument similar to the one needed in the present context was given in the appendix of \cite{peskin}.) This makes it very unlikely that the theory exhibits confinement without chiral symmetry breaking in this case, despite the existence of putative massless composite fermions that match the 't Hooft anomalies. This is because if $\tilde{\chi}$ were massless, then  $\psi\psi$ would be as well. Yet the only reason for the scalar $\psi\psi$ to be massless is if it is a Goldstone mode for broken chiral symmetry. 
 
\subsubsection*{A Comment on Symmetric Mass Generation on the Lattice}

In the context of putting chiral gauge theories on the lattice, the Weingarten inequalities bring a new hurdle. One may imagine using an auxiliary theory of the kind described in this paper to gap the fermion doublers, leaving behind a chiral gauge theory, along the lines first envisaged in \cite{ep}. However, if the auxiliary theory is itself involves chiral dynamics then the  problem has simply been shifted elsewhere: one would need yet an auxiliary auxiliary theory to gap the new doublers. The situation is potentially more straightforward when the auxiliary theory is vector-like. This, however, is when the Weingarten inequalities bite. 

There are, as we've seen, ways around this. By introducing additional scalar fields one evade the Weingarten inequalities and, as we've seen, exhibit the kind of symmetric mass generation required to lift the doublers. But this now comes with a new hurdle: the sign problem. This, it appears, will be a necessarily evil in any attempt to put a chiral gauge theory on the lattice.

\subsubsection*{Again, The Gapped Phase At Weak Coupling}

A weakly coupled description of the gapped phase   can be found with just a minor tweak from the previous example. First note that it is not sufficient to add extra scalar fields and Higgs $G\times H = SU(N) \times SU(N-4)$. The reason for this is obvious: even in the colour-flavour locked phase, it is not possible to preserve an $SU(N)$ global symmetry for the simple reason that $G\not\subset H$.

This also gives the clue for how to proceed. We start by augmenting our original fermions \eqn{g1} with further fermions, transforming in a vector-like representation of $G$. Specifically, we add four further $\overline{\Box} + \Box$ pairs. This is within the rules spelled out at the beginning of this section and,  because they are vectorlike, it is trivial to give them a mass and decouple them. However, we can subsequently  deform the theory so that they become massless. The new content of massless fermions then transforms as
\be G = SU(N)\ \mbox{with  $\anti$ and $N$  $\bar{\Box}$ and $4$ $\Box$}\nn\ee
Similar matter content was first discussed in \cite{by}. The theory now has the enhanced symmetry
\be \tilde{H}=SU(N)\nn\ee
As in previous examples, we add additional fermions that are singlets under $G$. We will need $\frac{1}{2} N(N+7)$ of these; their transformations under $\tilde{H}$ are listed in the table below where they appear as $\lambda$ and $\tilde{\rho}$. Finally, we add a scalar field $\phi$ whose transformation properties are also listed in the table.

The presence of these extra fermions gives rise to additional symmetries, so the full symmetry structure is now
\be G \times \tilde{H} \times SU(4) \times U(1)  = SU(N) \times SU(N)\times SU(4) \times U(1)\label{gh}\ee
Here we list only the unique  $U(1)$  that has no mixed anomaly with any non-Abelian factor.
Neither the $SU(4)$ nor $U(1)$ will play an important role in our story, but we keep track of them for completeness. The full collection of fermions and scalars now transform as
\FloatBarrier
\begin{table}[h!]
\begin{center}
\begin{tabular}{c|c|c|c|c|c||c}
& $ \chi$ & $\psi$ & $\lambda$ & $\rho$ &  $\tilde{\rho}$ & $\phi$ \\
  \hline 
 $G=SU(N)$ & $\anti$ & $\overline{\Box}$  & ${\bf 1}$ & $\Box$ & ${\bf 1}$ & $\Box$ \\
  $\tilde{H}=SU(N)$ & ${\bf 1}$ & $\overline{\Box}$  & $\sym$  & ${\bf 1}$ & $\overline{\Box}$ & $\overline{\Box}$ \\
    $SU(4)$ & ${\bf 1}$ & ${\bf 1}$  & ${\bf 1}$  & ${\bf 4}$ & $\bar{\bf 4}$ & ${\bf 1}$ \\    
   $U(1)$ & $-(N-4)$ & $N-2$  & $-N$  & $-(N-2)$ & $N$ & 2 \\    \end{tabular}
 \end{center}
\end{table}
\FloatBarrier

\noindent
Note that the $\psi$ fields in this table include the original chiral fermions, together with the 4 additional  fermions in the $\Box$ of $G$ that we added; these are compensated by the $\rho$ fermions that transform in the $\overline{\Box}$ of $G$.

The fermions are coupled through the Yukawa interactions 
\be {\cal L} = \phi\lambda\psi + \phi^\dagger \chi\psi + \phi^\dagger\tilde{\rho}\rho + {\rm h.c.}\nn\ee
Each of these terms is invariant under the full symmetry group \eqn{gh}. 
From here on, things are straightforward. A colour-flavour locked expectation value $\phi^a_{\ i} = v\delta^a_{\ i}$ breaks
\be G\times \tilde{H} \times U(1) \ \longrightarrow \ SU(N)_{\rm diag}\nn\ee
This time, the Abelian $U(1)$ does not survive. The  $SU(4)$, meanwhile, is unharmed. Under the surviving $SU(N)_{\rm diag}\times SU(4)$, the various fermions decompose as 
\be 
\chi:\ \ & \ \  (\anti,{\bf 1},{\bf 1}) \ &\longrightarrow\ \  \  (\anti,{\bf 1},{\bf 1}) \nn\\ 
\psi:\ \ &\ \ \ (\overline{\Box},\overline{\Box},{\bf 1}) \ \ &\longrightarrow\ \ \ (\overline{\sym},{\bf 1})\ \oplus\ (\overline{\anti},{\bf 1})\nn\\ 
\lambda:\ \  &\ ({\bf 1},\sym,{\bf 1}) \  &\longrightarrow \ \ \  (\sym,{\bf 1}) \nn\\
\rho:\ \  &\ (\Box,{\bf 1},{\bf 4}) \  &\longrightarrow \ \ \  (\Box,{\bf 4})\nn\\
\tilde{\lambda}:\ \  &\ ({\bf 1},\overline{\Box},\overline{\bf 4}) \  &\longrightarrow \ \ \  (\overline{\Box},\overline{\bf 4})\nn\ee
Clearly the fermions now sit in a vector-like representation of $SU(N)_{\rm diag}\times SU(4)$. It's simple to check that the Yukawa terms succeed in gapping all of them, preserving this symmetry.

\subsection{Standard Model Fermions}

In Nature, all fermions get a mass through the Higgs mechanism, with accompanying breaking of the gauge symmetry
\be G = SU(3)\times SU(2)_L \times U(1)_L\nn\ee
(The meaning of the unfamiliar $L$ subscript on the hypercharge gauge group will become apparent shortly.)

Nonetheless, it is interesting to ask whether there exists a different phase, in which all fermions are massive with $G$ unbroken. This may be useful to describe a fourth generation with mass significantly higher than the weak scale. Alternatively, as stressed in the introduction, it may offer a way to evade the Nielsen-Ninomiya theorem  \cite{nn1,nn2} by gapping the lattice doublers leaving  $G$ untouched. 

An example of the symmetric gapped phase was presented in \cite{shlomo}. There it was shown how one could gap a single generation of the Standard Model by invoking the kind of supersymmetric s-confinement ideas that we described in the previous section\footnote{For discussions of various discrete discrete obstacles to gapping Standard Model fermions, see \cite{garcia,nak,juv1,juv2}.}.

Here we ask: is there a weakly coupled description of this gapped phase? We will see that the answer is yes and, moreover, it appears to be closely related to certain extensions of the Standard Model that have been discussed in a more phenomenological context.

A single generation of the Standard model consists of a collection of 15 right-handed fermions, with quantum numbers

\FloatBarrier

\begin{table}[h!]
\begin{center}
\begin{tabular}{c|c|c|c|c|c}
& $ l$ & $q$ & $e$ & $u$ & $d$ \\
  \hline 
 $G$ &$({\bf 1},{\bf 2})_{-3}$ & $(\bar{\bf 3},{\bf 2})_{+1}$ & $({\bf 1},{\bf 1})_{+6}$ &  $({\bf 3},{\bf 1})_{-4}$ & $({\bf 3},{\bf 1})_{+2}$
  \end{tabular}
 \end{center}
\end{table}
\FloatBarrier

\noindent
In addition, we could add a 16$^{\rm th}$ right-handed neutrino, uncharged under $G$.

In common with the previous example (and, indeed, with \cite{shlomo}) we start by adding further fermions that are vector-like under $G$. These additional fermions will enhance the symmetries symmetries  beyond $G$, offering opportunities for introducing further gauge fields. Rather than building up from the Standard Model, we instead start by presenting the full theory. We will then explain how one can deform this theory in two ways: firstly, to reduce to the massless Standard Model fermions charged under $G$, and secondly to gap everything preserving $G$. In this way, we will demonstrate a path from massless, chiral fermions to a gapped symmetric phase.

Our parent theory is a parity-symmetric extension of the Standard Model, based on the group
\be K=G\times H\nn\ee
(omitting discrete quotients) where 
\be H=SU(2)_R\times U(1)_R\nn\ee
In contrast to previous sections, it will be useful to split fermions into a left-handed set and a right-handed set. 
The fermions and their quantum numbers under $K$ are

\FloatBarrier

\begin{table}[h!]
\begin{center}
\begin{tabular}{c|c}
{\rm left-handed} & {\rm right-handed}\\ \hline 
$q: \ ({\bf 3};{\bf 2},{\bf 1})_{-1,+2}$ & $d:\ ({\bf 3};{\bf 1},{\bf 2})_{+2,-1}$\\
$d': \ ({\bf 3};{\bf 1},{\bf 1})_{+2,-4}$ & $u:\ ({\bf 3};{\bf 1},{\bf 1})_{-4,+2}$\\
$l': \ ({\bf 1};{\bf 2},{\bf 1})_{-3,0\ }$ & $\nu:\ ({\bf 1};{\bf 1},{\bf 2})_{0,-3\ \ }$\\
$l: \ ({\bf 1};{\bf 2},{\bf 2})_{+3,-3}$ & $e:\ ({\bf 1};{\bf 1},{\bf 1})_{+6,-6\ }$\\
\end{tabular}
 \end{center}
\end{table}
\FloatBarrier

\noindent
Written in this way, it is clear that the theory is parity invariant. Parity acts by exchanging left-handed and right-handed spinors while, at the same time, swapping $SU(2)_L\leftrightarrow SU(2)_R$ and $U(1)_L\leftrightarrow U(1)_R$.

 Although the fermion content above is parity invariant, it remains  chiral and no quadratic mass is possible preserving $K$ (or, indeed, the subgroup $G$). It can be checked that $K$ is anomaly free.
 
 Left-right symmetric extensions of the Standard Model have long been proposed as possible theories of the world, starting with the work of Pati and Salam \cite{ps} and followed by many variants \cite{mohap1,mohap2,senaj,barr}\footnote{It is curious that the Pati-Salam model recently made an appearance in symmetric mass generation in the context of reduced staggered (or K\"ahler-Dirac) fermions on the lattice \cite{simon2}. It would be interesting to understand the connection between these related set-ups.}. The gauge group and matter content above differs from these set-ups, although it does coincide with the reduction of trinification models\footnote{I'm grateful to Joe Davighi for pointing this out to me.}
 based on the gauge group $SU(3)^3$. (The original trinification papers  appear only in proceedings \cite{trin1,trin2}; the first readily available paper discussing this theory is \cite{trin3}.)

The  same matter content was also introduced in \cite{shlomo}, although the parity-invariant nature of the theory wasn't mentioned. In that paper, the fermions were gapped without breaking $G$ by relying on the strong coupling dynamics of $SU(2)_R\subset H$ (after a suitable supersymmetrisation.) Here, instead, we use a Higgs mechanism.

First, let's see how these fermions reduce to the familiar fermions in one generation of the Standard Model. We introduce two Higgs fields, with quantum numbers
\be h_L:  \ ({\bf 1};{\bf 2},{\bf 1})_{-3,0}\ \ \ \ {\rm and}\ \ \ \ h_R:\ ({\bf 1};{\bf 1},{\bf 2})_{0,-3}\nn\ee
Here $h_L$ carries the quantum numbers under $G$ of the Standard Model Higgs. Obviously $h_R$ is its parity conjugate. To reduce to the Standard Model we need both Yukawa terms
\be {\cal L}_{\rm Yuk} &=& h_L u^\dagger q + h_L^\dagger l^\dagger \nu + h_Rd^{\prime \dagger} d + h_R^\dagger l^{\prime }l +  {\rm h.c.}\label{smyuks}\ee
and dimension five operators
\be {\cal L}_{\rm dim-5} =  (h_L^\dagger l')(h_L^\dagger l') +  (h_R^\dagger \nu)(h_R^\dagger \nu) + h_L h_R^\dagger \Big[q^\dagger d +  l^\dagger e +  l^{\prime \dagger}\nu\Big] + {\rm h.c.} \label{dim5}\ee
both of which are  invariant under the full symmetry group $K$.

First  consider what happens when $h_R\neq 0$ with $h_L=0$.  This leaves $G$ unbroken, but breaks 
\be H = SU(2)_R \times U(1)_R \rightarrow U(1)_{B-L}\nn\ee
 The Yukawa interactions above give a mass to $d'$ and $l'$, each of which pairs up with  one-half of the $SU(2)_R$ doublets $d$ and $l$ respectively. The dimension 5 term $(h_R^\dagger \nu)(h_R^\dagger \nu)$  gives a mass to half of the $SU(2)_R$ doublet $\nu$. The theory above then reduces to the Standard Model with 16 fermions comprising one generation. As the name suggests, the action of the  unbroken $U(1)_{B-L}\subset H$ on the massless fermions coincides with $B-L$ symmetry of the Standard Model. (An additional scalar field must be introduced to gap the $U(1)_{B-L}$ gauge boson.)

With $h_R\neq 0$, the remaining terms in \eqn{smyuks} and \eqn{dim5} reduce to the Yukawa terms (and dimension 5 Majorana masses) of the Standard Model. Curiously,  the down-sector and electron Yukawa couplings in the Standard Model descend from dimension 5 operators.

We now turn to the main question: can we gap all fermions preserving the Standard Model group $G$? For this we must introduce a new scalar field carrying quantum numbers under $K$
\be \phi:\ ({\bf 1};{\bf 2},{\bf 2})_{-3,+3}\nn\ee
We couple this to the fermions above as
\be {\cal L} = \phi q^\dagger d + \phi l^{\prime \dagger}\nu + \phi^\dagger l^\dagger e + \phi^2 u^\dagger d' + {\rm h.c.}\label{smsurvive}\ee
An expectation value for $\phi$ breaks
\be K\ \longrightarrow\ G' \times U(1)_{B-L} =[SU(3) \times SU(2)_{\rm diag}\times U(1)_{\rm diag}]\times U(1)_{B-L}\label{survive}\ee
It is natural in this context to take all of $K$ to be gauged. Alternatively, to align more with previous examples, we could view $G= SU(3)\times SU(2)_L\times U(1)_L$ to be a global symmetry  and $H=SU(2)_R\times U(1)_R$ to be a gauge symmetry (a choice which obviously breaks parity) then the surviving symmetry group $G'$ is a global symmetry and should be viewed as  $G$ twisted with the gauge symmetry. 

It is simple to check that the quantum numbers of the fermions under this surviving symmetry rearrange themselves to become vector-like:
\FloatBarrier

\begin{table}[h!]
\begin{center}
\begin{tabular}{c|c}
{\rm left-handed} & {\rm right-handed}\\ \hline 
$q: \ ({\bf 3},{\bf 2})_{+1}$ & $d:\ ({\bf 3},{\bf 2})_{+1}$\\
$d': \ ({\bf 3},{\bf 1})_{-2}$ & $u:\ ({\bf 3},{\bf 1})_{-2}$\\
$l': \ ({\bf 1},{\bf 2})_{-3}$ & $\nu:\ ({\bf 1},{\bf 2})_{-3}$\\
$l: \ ({\bf 1},{\bf 1})_{0}$ & $e:\ ({\bf 1},{\bf 1})_{0}$\\
\end{tabular}
 \end{center}
\end{table}
\FloatBarrier

\noindent
Furthermore,  the Yukawa interactions and dimension 5 operator in  \eqn{smsurvive} succeed in gapping all the fermions, preserving the (now twisted)  symmetry $G$. Note that, in contrast to the strong coupling mechanism of \cite{shlomo}, the weak coupling approach describes a phase in which 16 fermions are gapped preserving $U(1)_{B-L}$ in addition to $G$.

As in previous examples, we  can't take a path from the gapless fermions to the gapped phase without passing through strong coupling. Here, this occurs because a simultaneous expectation value for both $h_R$ or $h_L$, together with $\phi$, would break $G$. We must turn $h_L$  and $h_R$ off before $\phi$  turns on. But when both are turned off, the $SU(2)_R$ gauge group flows to strong coupling.

\section{Gapping Chiral Fermions in Two Dimensions}\label{2dsec}

The story of symmetric mass generation in two dimensions differs from its 4d counterpart in a number of ways, not least the because of the Coleman-Mermin-Wagner theorem which prohibits the spontaneous breaking of continuous symmetries \cite{mw,coleman}.

Given the impossibility of spontaneous symmetry breaking, one might naively think that we could simply add Yukawa terms and condense the scalars to give the fermions a mass, safe in the knowledge that infra-red effects will ultimately restore the symmetry. However, such a process typically results in interacting gapless modes \cite{witten,affleck}. For  theories in which one can identify a large $N$ limit, these gapless modes can be viewed as Goldstone bosons but otherwise they form a CFT.  To achieve symmetric mass generation, one must find a way to gap these would-be Goldstone modes. As we will see in examples below, this is largely a question of finding appropriate relevant operators.

In this section we describe a number of different examples of symmetric mass generation in 2d, with results for Abelian and non-Abelian continuous symmetries in Sections \ref{absec} and \ref{nonabsec} respectively.

\subsection{Abelian Symmetries}\label{absec}

We start by considering chiral fermions transforming under $U(1)$ symmetries.  There is a long literature concerning possible gapping mechanisms for such fermions, with initial discussions focussed on the edge modes of quantum Hall states \cite{haldane}.  Much of this literature has in mind an underlying lattice, and the relevance or irrelevance of various gapping operators is not of much concern. Here, instead we focus on mechanisms for gapping in the continuum\footnote{There is a close connection between gapped phases and boundary states preserving $U(1)$ symmetries, and the latter have been studied in some detail in recent years \cite{kristen,ryu,bs1,bs2,bs3,thorngren,simeon}. (See, also \cite{sagi,juan} for earlier work on such states.) More recently, an interesting connection to zeros of Greens functions has been mooted \cite{latest}.}.

Instead of presenting the discussion in full generality, we will instead focus on a simple example that highlights the main ideas. We consider  chiral fermions transforming under a $G=U(1)$ symmetry. We take  two left-moving fermions $\psi_1$ and $\psi_2$  and two right-handed Weyl fermions $\tilde{\psi}_1$ and $\tilde{\psi}_2$,  transforming under the global symmetry $G=U(1)$ with charges:
\FloatBarrier

\begin{table}[h!]
\begin{center}
\begin{tabular}{c|c|c||c|c}
& $ \psi_1$ & $\psi_2$ & $\tilde{\psi}_1$ & $\tilde{\psi}_2$  \\
  \hline 
 $G$ & 3 & 4  &5 & 0  \end{tabular}
 \end{center}
\end{table}
\FloatBarrier

\vspace{-5mm}
\noindent
The gravitational anomaly vanishes as $c_L= c_R$, while the 't Hooft anomaly for $G$ is also vanishing since $3^2+4^2 = 5^2$. This means that it should be possible to gap the system without breaking $G$. The question is: how? As we'll see, there are a number of related ways to achieve this. 

\subsubsection*{Gapping with Multi-Fermion Terms}\label{wwsec}

The first mechanism follows ideas  of \cite{haldane} and has been emphasised in the context of chiral lattice gauge theories by Wang and Wen \cite{request,ww1}. (See also \cite{poppitz2} for a closely related approach.) Here, one introduces the following six-fermion operators
\be {\cal L} = g_1\,(\psi_1^\dagger)^2 \psi_2^\dagger \tilde{\psi}_1^2\tilde{\psi}_2^\dagger + g_2\,\psi_1^\dagger \psi_2^2 \tilde{\psi}_3^\dagger (\tilde{\psi}_2^\dagger)^2 + {\rm h.c.}\label{haldane}\ee
Here the quadratic terms are to be viewed through point-splitting regularisation and hide a  derivative: $\psi_1^2 = \psi_1\partial_-\psi_1$. 
Both terms are invariant under both Lorentz symmetry and $G$. However, both terms are irrelevant: they are dimension 5 operators.

Nonetheless, in $d=1+1$ dimensions, with Abelian symmetries, there is a way forward. That is because theories of free fermions in $d=1+1$ dimensions sit at a special point in a moduli space of conformal theories. This is the Narain moduli space which, for $N$ Dirac fermions, has the form
\be {\cal M}_N = \frac{O(N,N,{\bf R})}{O(N,{\bf R})\times O(N\times {\bf R})\times O(N,N,{\bf Z})}\nn\ee
This means that these theories admit ${\rm dim}\,{\cal M}_N = N^2$ exactly marginal operators. These are simply current-current interactions where, in the present case, the currents involved are associated to the $G=U(1)$ chiral symmetry and an orthogonal $H=U(1)$ symmetry that we describe below. The exact details will not be needed for our story but, suffice to say, one can find regions in ${\cal M}_N$, sufficiently far from the free fermion point,  where the two operators in \eqn{haldane} become relevant. This means that the interaction \eqn{haldane}, together with suitable marginal deformations, can gap the system preserving $G$ in the continuum, as well as on the lattice.

\subsubsection*{Gapping with a Gauge Interaction}

There is an alternative, but related, method to gap fermions preserving the chiral symmetry $G$, one that is similar to the approaches that we  used in four dimensions. This time, we first observe that, in addition to $G$, the system admits a second non-anomalous chiral symmetry, $H=U(1)$, with charges
\FloatBarrier
\begin{table}[h!]
\begin{center}
\begin{tabular}{c|c|c||c|c}
& $ \psi_1$ & $\psi_2$ & $\tilde{\psi}_1$ & $\tilde{\psi}_2$  \\
  \hline 
 $G$ & 3 & 4  &5 & 0  \\
  $H$ & 5 & 0  &3 & -4  \end{tabular}
 \end{center}
\end{table}
\FloatBarrier
\noindent
Importantly, there are no mixed anomalies between $G$ and $H$. 

To gap the fermions without breaking $G$, we first introduce new, heavy degrees of freedom. These are gauge bosons for $H$, Higgsed at a high scale by a scalar $\phi$ carrying charge $(G,H)=(0,1)$. The 2d gauge coupling $e^2$ sets a scale and, as  the mass $m_\phi$ of the scalar varies, the theory has two different phases:
\begin{itemize}
\item $m_\phi^2 \ll -e^2$: In this phase, the $H=U(1)$ gauge field is Higgsed and massive. The massless spectrum consists of the original fermions $\psi_i$ and $\tilde{\psi}_i$, with any interactions suppressed by $e^2/|m_\phi|^2$. In particular, as $m_\phi^2/e^2\rightarrow -\infty$ and the extra degrees of freedom decouple, we return to the original free fermions. 

\item $m_\phi^2 \gg +e^2$: In this regime, the scalar field decouples and the $H=U(1)$ gauge field confines. At low-energies, we now have just a single massless Dirac fermion, consisting of a left-mover $\Psi$ and a right-mover $\tilde{\Psi}$. These are gauge invariant states given by
\be \Psi = \psi_2\ \ \ {\rm and}\ \ \ \ \tilde{\Psi} =\frac{1}{e^2} \psi_1^2(\tilde{\psi}_1^\dagger)^2 \tilde{\psi}_2\nn\ee
Note that the charge under the original $G=U(1)$ symmetry is $G[\Psi]= 4$ and $G[\tilde{\Psi}] = -4$.
We now add the six-fermion operator in the UV
\be {\cal L}_{UV} = g\,\psi_1^2\psi_2(\tilde{\psi}_1^\dagger)^2 \tilde{\psi}_2\label{hal2}\ee
This is invariant under $G$. After confinement, this will descend to a mass term 
\be {\cal L}_{IR} = ge^2\, \Psi\tilde{\Psi}\label{gmass}\ee
This gaps the fermions, preserving the symmetry $G$.
\end{itemize}
Although the details vary,  the key aspect of this gauge theoretic gapping mechanism are the same as the multi-fermion terms described previously. Indeed, the operator \eqn{hal2} coincides with the first term in the multi-fermion couplings \eqn{haldane}. 

To further highlight the connection between the two approaches, in the phase $m_\phi^2\ll -e^2$  where we have massless fermions, the heavy photon induces current-current interactions of the form
\be {\cal L}_{\rm marginal} \sim \frac{e^2}{|m_\phi^2|}\, J^\mu J_\mu\label{eddiedog}\ee
This means that, in the Higgs phase, the massless fields don't sit at the free fermion point, except at $m^2_\phi/e^2\rightarrow -\infty$. Instead the gauge interactions induce the marginal coupling  \eqn{eddiedog} and  we have an interacting CFT in the infra-red. As we vary the dimensionless ratio $m^2_\phi/e^2$, we trace a path in the moduli space ${\cal M}_N$. As we do so the operator \eqn{hal2} changes its dimension. By the time we reach the confining phase, this operator has become relevant, as clearly seen when written in its gauge invariant form \eqn{gmass}.

\subsubsection*{The Gapped Phase at Weak Coupling}\label{pqsec}

The gauge theory description above gives a straightforward path from the free fermion phase to the gapped phase. However, it clearly relies on the strong coupling physics of confinement in the final stage. It is not hard to find a weakly coupled description of this gapped phase.

In fact, there are a number of ways distinct ways to achieve this, each introducing a scalar field with different charges under $(G,H)$. As one example, we introduce a scalar field $p$ with charge $(G,H) = (-3,-1)$, and the Yukawa interactions
\be {\cal L} =\lambda\left(p \psi_1\tilde{\psi}_2 + p^3\psi_2\tilde{\psi}_1 + {\rm h.c.}\right)
\label{yuk3}\ee
where $\lambda$ is a mass scale.  Note that the second term isn't quite of Yukawa form since it involves $p^3$ rather than just a single scalar field. However, in $d=1+1$ dimensions, scalar fields are dimensionless so this term has the same dimension as the genuine Yukawa term.

Both terms are invariant under both the global symmetry $G$ and the gauge symmetry $H$. Our goal, as always, is to gap fermions preserving the symmetry $G$. We can do this by giving an expectation value to $p$. Of course, since $p$ carries charge under $G$, it naively looks as if the global symmetry  is broken. But it survives by twisting the gauge symmetry, and the surviving global symmetry is $G' = G-3H$. Importantly, the chiral quantum numbers under $G$ morph into a vector-like quantum numbers under $G'$; this, as we've seen in previous examples, is only possible because the original symmetry was anomaly free. 
When $p$ gets an expectation value, all fermions get a mass through the interaction \eqn{yuk1}, with $G'$ unbroken.
 

\subsection{Non-Abelian Symmetries}\label{nonabsec}

We now turn to a class of $d=1+1$ dimensional theories that realise non-Abelian symmetries in a chiral manner. These can be viewed as 2d counterparts to the 4d theories described in Section \ref{4dsec} and exhibit many similar features.

In Section \ref{4dsec}, we first introduced the massless, chiral fermions, charged under a global symmetry $G$. We then found an auxiliary gauge theory, with gauge group $H$ which could be corralled into providing a mechanism for symmetric mass generation. In that case, the strong coupling dynamics of $H$ was well known, either from studies of supersymmetric theories \cite{is,ip} or of chiral theories \cite{raby}. Here we reverse the presentation. 

In two dimensions, there do not seem to be examples of strongly coupled supersymmetric gauge theories in the literature that can be put to work for symmetric mass generation. (The Seiberg-like ${\cal N}=(2,2)$ dualities of \cite{hori1,hori2} don't have the chiral symmetries necessary to do the job, while the interesting ${\cal N}=(0,2)$ dualities of \cite{razamat,sachi} have free fields in the infra-red whose UV origin is difficult to determine.)

Nonetheless, it is not hard to construct non-supersymmetric gauge theories that will do the job.   Below we describe a pair of strongly interacting 2d gauge theories and offer conjectures for their infra-red behaviour. We will then see how these can be employed for symmetric mass generation.

\subsubsection*{\underline{$H=SU(r)$ with an Anti-symmetric and Fundamentals}}

As in Section \ref{4dsec}, we will refer to global symmetry groups as $G$ and gauge groups as $H$. We further use the convention that right-moving fermions come with a tilde above, while left-moving fermions do not.

Our first example is an $H=SU(r)$ gauge theory, with a $G=SU(r-2)$ global symmetry. In addition, there is a $U(1)_L\times U(1)_R$ global symmetry so that under the combination $H\times G\times U(1)^2$, the fermions transform as
\be \mbox{left-mover $\psi$:} && (\overline{\Box},{\Box})_{1,0}\nn\\
\mbox{right-mover $\tilde{\chi}$:} && (\anti,{\bf 1})_{0,1} \nn\ee
The theory has no gauge anomaly\footnote{In 2d, non-Abelian anomalies are associated to the Dynkin index $\mu$ of the representation. In what follows, we will need that, for $H=SU(r)$, $\mu(\Box) =1$ and $\mu(\sym) = r+2 $ and $\mu(\anti) = r-2$.}  for $H$. 
What is the low-energy physics? There are a number of simple but important restrictions. 

First, the theory has a gravitational anomaly, with $c_L-c_R=\frac{1}{2}(r-2)(r-1)-1>0$, and so must, at least, have massless left-moving modes. 

Furthermore, there are anomalies for the global symmetries, each of which descends to the level of  a current algebra in the infra-red
\be &&{\rm Tr}\,\gamma^3 \,SU(r-2)^2 =  r\label{anom1}\\  && {\rm Tr}\,\gamma^3\, U(1)_L^2= r(r-2)\label{anom2}\\ && {\rm Tr}\,\gamma^3 \, U(1)_R^2 = -\frac{1}{2}r(r-1)\label{anom3} \\  &&{\rm Tr}\,\gamma^3 \,U(1)_R\cdot U(1)_L = 0 \label{anom4}\ee
These must all be reproduced at low energy.

In fact this there is a natural candidate for the infra-red phase as  a confined theory of free, massless, gauge invariant fermions. In the left-moving sector, the simplest gauge invariant fermion is
\be \lambda = \psi\tilde{\chi}\psi\nn\ee
This alone does not replicate the anomalies. Indeed, the gravitational anomaly shows that it must be accompanied by a single right-moving partner. It is more complicated to construct a gauge invariant, right-moving fermion: the simplest is the dibaryonic object
\be \tilde{\cal B} = \epsilon^{i_1\ldots i_{r-2}} \,\psi_{a_1i_1}\ldots \psi_{a_{r-2}i_{r-2}} \,\epsilon_{b_1\ldots b_r}\,\tilde{\chi}^{a_1b_1}\ldots\tilde{\chi}^{a_{r-2}b_{r-2}}\tilde{\chi}^{b_{r-1}b_r}\label{rhotilde}\ee
where the $i=1,\ldots, r-2$ index is associated to the $G=SU(r-2)$ flavour group, and the $a,b=1,\ldots,r$ index to the $H=SU(r)$ gauge group. In the UV, the fermionic operator $\tilde{\cal B}$ has dimension $r-\frac{3}{2}$. Nonetheless, we conjecture that both $\tilde{\cal B}$ and $\lambda$  flow to a free, massless fermions in the infra-red. Indeed, it is not unusual to find massless baryons arising in 2d non-Abelian gauge theories \cite{baluni,inflationrocks,rabi}. Under the $SU(r-2)\times U(1)_L \times U(1)_R$  symmetry, these two fermions have quantum numbers
\be \lambda: \sym_{2,1}\ \ \ \ {\rm and}\ \ \ \tilde{\cal B}:\ {\bf 1}_{r-2,r-1}\nn\ee
It is straightforward to check that these reproduce the anomalies \eqn{anom1}-\eqn{anom4}.

\subsubsection*{\underline{$H=SU(r)$ with a Symmetric and Fundamentals}}

We also consider the counterpart non-anomalous chiral theory with $H=SU(r)$ gauge group and $G=SU(r+2)$ global symmetry. Under $H\times G \times U(1)_L\times U(1)_R$, the fermions now transform as
\be \mbox{left-mover $\psi$:} && (\overline{\Box},{\Box})_{1,0}\nn\\
\mbox{right-mover $\tilde{\lambda}$:} && (\sym,{\bf 1})_{0,1} \nn\ee
This time $c_L-c_R=\frac{1}{2}(r+2)(r+1)-1>0$ suggesting that the infra-red may consist of a  left-moving fermion in the $\anti$ of $G=SU(r+2)$, together with a single right-mover. The obvious candidate for the former is 
\be \chi = \psi\tilde{\lambda}\psi\nn\ee
This transforms under $G\times U(1)_L\times U(1)_R$ as $\overline{\anti}_{2,1}$. The putative right-mover  should be a baryonic state of the form $\tilde{\cal B}' \sim \psi^{r+2}\tilde{\lambda}^{r+1}$, which transforms as ${\bf 1}_{r+2,r+1}$. To form this state we take
\be \tilde{\cal B}' = \epsilon^{i_1\ldots i_{r+2}}\,\psi_{i_1a_1}\ldots \psi_{i_ra_r}\psi_{i_{r+1}c}\psi_{i_{r+2}d} 
\,\epsilon_{b_1\ldots b_r}\tilde{\lambda}^{a_1b_1}\ldots \tilde{\lambda}^{a_rb_r}\partial_-\tilde{\lambda}^{cd}\label{baryon2}\ee
where the $\partial_-$ is necessary to ensure that $\tilde{B}$ doesn't vanish on symmetry grounds. Moreover, it makes $\tilde{B}$ is a right-moving fermion even though its constituents include one more left-mover  $\psi$ than right-movers $\tilde{\lambda}$. 
Once again, one can check that $\chi$ and $\tilde{\cal B}$ saturate all the 't Hooft anomalies of this model.

\subsubsection*{Symmetric Mass Generation}

Under the assumption that the theories  described above do indeed confine as claimed, then it is straightforward to employ them for symmetric mass generation. To this end, we start with a collection of fermions transforming under 
\be K= SU(N) \times SU(N+2) \times U(1)_L\times U(1)_R\nn\ee
with quantum numbers
\be \mbox{left-movers:}&&\ \ \ \left\{\begin{array}{l} \psi:\ \  ({\Box},\overline{\Box})_{1,0} \\ \rho:\ \ ({\bf 1},{\bf 1})_{N+2,N+1}
\end{array}\right. \nn\\\mbox{right-movers:}&&\ \ \  \left\{\begin{array}{l} \tilde{\chi}:\ \ ({\bf 1},\anti)_{-2,-1} \\
 \tilde{\lambda}:\ \ (\sym,{\bf 1})_{0,1}
\end{array}\right. \nn\ee
These fermions have $c_L=c_R = (N+1)^2$ and all anomalies of $K$ vanish.

We now gauge a subgroup of $K$. Either we pick $G=SU(N)$ as a global symmetry and gauge $H=SU(N+2)$, or vice versa. As in the previous section, we denote the gauge coupling by $e^2$. In either case, we also introduce scalar fields $\phi$, charged only under $H$ which have the effect of Higgsing the gauge group. Now the low-energy physics is expected to depend on  $m_\phi^2$ as follows:
\begin{itemize}
\item $m_\phi^2 \gg e^2$: In this regime, the scalar field decouples and the $H$ gauge theory confines, leaving behind gapless degrees of freedom which, we assume, are the massless fermions described previously. These fermions can now be gapped in the usual fashion by the addition of dangerously irrelevant multi-fermion terms. For $H=SU(N)$, these  schematically take the form
\be {\cal L}_{H=SU(N)} =  \tilde{\lambda}^\dagger\psi\tilde{\chi}\psi  + \rho^\dagger\psi^N\tilde{\chi}^{N+1}\nn\ee
where the second term involves the baryon \eqn{rhotilde}. For $H=SU(N+2)$, these take the form
\be {\cal L}_{H=SU(N+2)} =  \tilde{\lambda}^\dagger\psi\tilde{\chi}\psi + \rho^\dagger \psi^{N+2} \partial_-\tilde{\lambda}^{N+1}\nn\ee
In both cases, the fermions are gapped preserving $G\times U(1)^2$ as a global symmetry.
\item $m_\phi^2 \ll -e^2$: In this regime, the gauge field lives in the Higgs phase.  As in the $U(1)$ case, we can integrate out the gauge bosons to generate current-current interactions terms
\be L_{\rm non-Abelian} = \frac{e^2}{|m_\phi^2|} {\rm Tr}\,J_H^\mu J_{H\,\mu}\nn\ee
For the 2d $U(1)$ gauge theories described in the  previous section these current-current interactions are exactly marginal. That resulted in a Kosterlitz-Thouless phase transition at  $m^2_\phi \sim -e^2$  where the higher fermion terms became relevant. In contrast, non-Abelian current-current interactions are marginally relevant. This means that there is no phase transition between the Higgs phase and confining phase: the theory remains gapped for all finite $m_\phi^2$. 
\end{itemize}
The discussion above turns the question of symmetric mass generation on its head. If one wants to use non-Abelian 2d gauge dynamics to drive symmetric mass generation, then the difficulty is in finding the gapless phase rather than the gapped! In the set-up above, it arises only in the strict $m_\phi^2/e^2\rightarrow -\infty$ limit. Of course, this is simply the statement that 2d theories with non-Abelian symmetries have marginally relevant current-current interactions. In these examples, one can simply omit the gauge interactions completely and drive symmetric mass generation with these current interactions, now preserving the entire symmetry group $K$. In the general setting, the problem is to understand what current interactions will succeed in gapping the theory.


%
%



\acknowledgments
 Nathan Seiberg  raised  the possibility of gapping at weak coupling at a talk I gave in December 2020 \cite{mytalk}. I'm particularly grateful to him for this suggestion and for subsequent communications. I'd also like to thank Pietro Benetti Genolini,  Philip Boyle Smith, Joe Davighi, Avner Karasik, Nakarin Lohitsiri, Kaan Onder, Shlomo Razamat,  Matt Strassler, Carl Turner and Juven Wang for comments and  conversations on this topic, and Stefano Bolognesi, Ken Konishi and Andrea Luzio for discussions on related issues.
I'm supported by the STFC consolidated grant ST/P000681/1, a Wolfson Royal Society Research Merit Award holder and a Simons Investigator Award.

\end{document}